\documentclass[submission, Proceedings]{SciPost}

\hypersetup{
    colorlinks,
    linkcolor={red!50!black},
    citecolor={blue!50!black},
    urlcolor={blue!80!black}
}

\usepackage[bitstream-charter]{mathdesign}
\DeclareSymbolFont{usualmathcal}{OMS}{cmsy}{m}{n}
\DeclareSymbolFontAlphabet{\mathcal}{usualmathcal}

\begin{document}

\begin{center}{\Large \textbf{
Medium modification of jet shape observables in Pb-Pb collisions at $\sqrt{s_{NN}}$= 2.76 TeV using EPOS and JEWEL event generators
\\
}}\end{center}

\begin{center}
Sumit Kumar Saha\textsuperscript{1*},
Debojit Sarkar\textsuperscript{2},
Subhasis Chattopadhyay\textsuperscript{1},
Ashik Ikbal Sheikh\textsuperscript{1} and
Sidharth Kumar Prasad\textsuperscript{3}

\end{center}

\begin{center}
{\bf 1} Experimental High Energy Physics and Applications Group,HBNI, Variable Energy Cyclotron Center, 1/AF, BidhanNagar,  Kolkata-700 064, INDIA
\\
{\bf 2} Wayne State University, 666 W. Hancock, Detroit, MI 48201, USA
\\
{\bf 3} Bose Institute, 93/1, APC Road, Kolkata-700009
\\
* sumitkumarsaha92@gmail.com
\end{center}

\begin{center}
\today
\end{center}


\definecolor{palegray}{gray}{0.95}
\begin{center}
\colorbox{palegray}{
  \begin{tabular}{rr}
  \begin{minipage}{0.1\textwidth}
    \includegraphics[width=30mm]{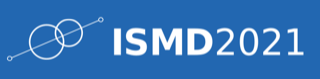}
  \end{minipage}
  &
  \begin{minipage}{0.75\textwidth}
    \begin{center}
    {\it 50th International Symposium on Multiparticle Dynamics}\\ {\it (ISMD2021)}\\
    {\it 12-16 July 2021} \\
    \doi{10.21468/SciPostPhysProc.?}\\
    \end{center}
  \end{minipage}
\end{tabular}
}
\end{center}

\section*{Abstract}
{\bf
The substructure of charged jets gets modified due to the presence of the medium in heavy-ion collisions which is an indication of redistribution of energy inside the jet cone. It helps to understand the energy loss mechanisms of jets in the medium. A model dependent study of the modification of differential jet shape $(\rho(r))$ and the angularity (g) due to medium, has been performed in the most central Pb-Pb collisions at 2.76 TeV in the jet range of 20-40 GeV/c. 
}


\section{Introduction}
\label{sec:intro}
A partonic medium consisting of quarks and gluons is formed in the collision of heavy ions at high energies ~\cite{r1}. Partons suffer energy loss through radiation or collision as they travel in the medium and their fragmentation functions get modified. The internal structure of jet also undergoes modification. Thus, measurement of such observables in central heavy ion collisions with a comparison to pp collisions can help to understand the jet-medium interaction ~\cite{r2}. 
The motivation is to study jet-shape observables at lower $p_{T}$ without considering the recoiled partons using JEWEL ~\cite{r3} which helps to understand the effect of the recoiled medium partons through a comparison to the experimental data and to study the same using EPOS-3 ~\cite{r4,r5,r6,r7} with a simplistic partonic energy loss mechanism and secondary hard-soft interactions.
\section{Observables and analysis method}
Two jet shape observables, differential jet-shape $\rho(r)$ ~\cite{r8} and angularity (g) or girth ~\cite{r9} have been studied. $\rho(r)$ describes the radial distribution of the jet $p_T$ density inside the jet cone and is defined as,
\begin{equation}
\rho(r)= \frac{1}{\delta r}\frac{1}{N_{jet}}\sum_{jets}\frac{\sum_{tracks\epsilon[r_a,r_b]} p_T^{track}}{p_T^{jet}}
\end{equation}
The jet cone is divided into several annuli with radial width of $\delta r$ and inner radius of $r_a$=r-$\delta r$/2 and outer radius of $r_b$=r+$\delta r$/2. r is the radial distance of the track from the jet axis and is defined as,
\begin{equation}
r=\sqrt{(\phi^{track}-\phi^{jet})^2+(\eta^{track}-\eta^{jet})^2} \le R
\end{equation}
where R is the jet radius.
The angularity is defined as,
\begin{equation}
g=\sum_{i \epsilon jet}\frac{p_T^i}{p_T^{jet}}|\Delta R^i_{jet}|
\end{equation}
where $\Delta R$ is the distance between i-th constituent and the jet axis in ($\eta,\phi$) space.
These shape observables describe the radial distribution of the jet energy inside the jet cone.
Charged jets reconstruction has been done with anti-$k_T$ jet finding algorithm using Fastjet package ~\cite{r10} with two resolution parameters R = 0.2 and R = 0.3 for 20$< p_{T,ch}^{jet} <$40 GeV/c ~\cite{our}. Tracks with $|\eta|<$0.9 and $p_{T,min}$=0.15 GeV/c are used for jet reconstruction and $|\eta_{jet}|<$0.7 and 0.6 for R = 0.2 and R = 0.3 are selected respectively. Background contribution is reduced by selecting jets having at least one particle with $p_T >$ 5 GeV/c ~\cite{r11,r12}.
We have compared the Pb-Pb results from both the models ~\cite{our} to the pp results obtained by JEWEL representing no medium effect.
\section{Results}
\begin{figure}[h]
	\includegraphics[width=0.41\textwidth,height=0.38\textwidth]{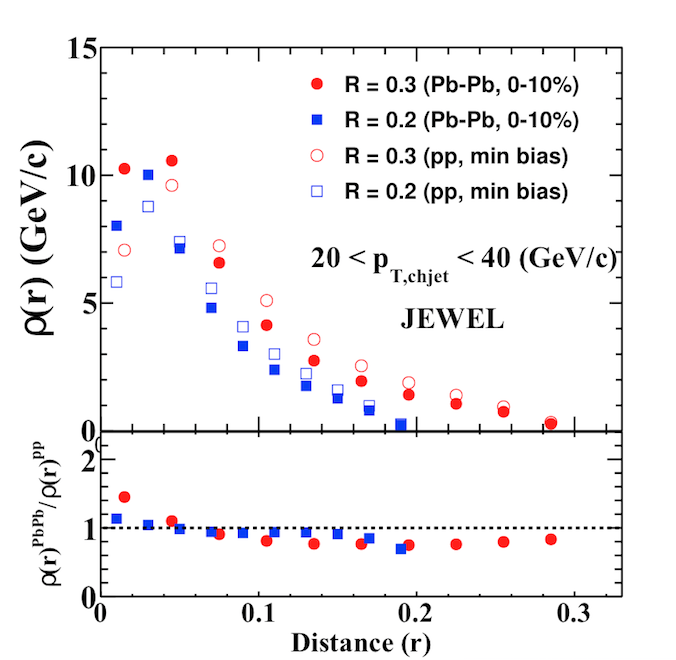}
	\hfill
	\includegraphics[width=0.4\textwidth,height=0.37\textwidth]{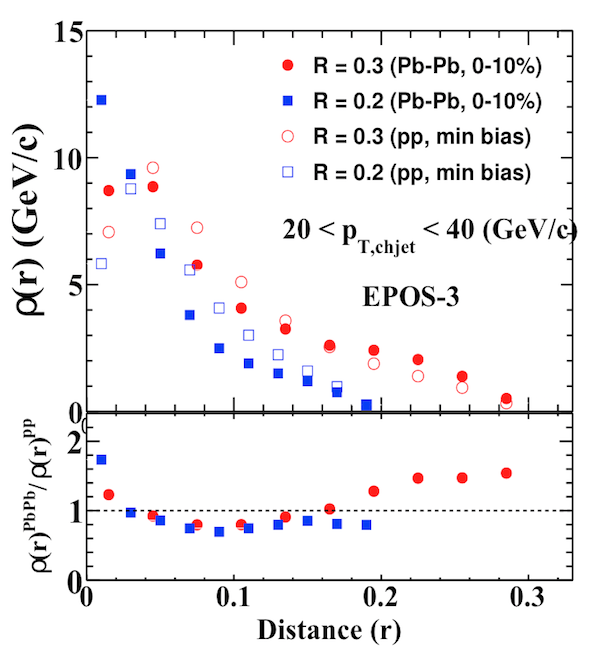}
	\caption{$\rho(r)$ as a function of distance from the jet axis for inclusive charged jets in Pb-Pb 0-10\% using the JEWEL (recoil OFF) and EPOS3 event generators compared with theminimum bias pp results along with their ratios ~\cite{our}}
	\label{ref1}
\end{figure}

As shown in Fig~\ref{ref1} ~\cite{our}, the ratio of $\rho(r)$ distribution of 0-10\% Pb-Pb to pp collisions deviates from unity which is an indication of a modification to the jet structure due to the presence of the medium. 
Increasing the R to 0.3 includes the energy carried away by softer particles at larger angles from the jet axis. At higher r, the ratio becomes $>$1 that indicates broadening of jets at the periphery in EPOS-3 which is qualitatively consistent with the experimental observations. The energy lost due to jet-medium interaction is distributed at larger distances from the jet axis and it signifies medium induced modification to the internal jet structure ~\cite{r8}.
 \begin{figure}[h]
 	\includegraphics[width=0.41\textwidth,height=0.37\textwidth]{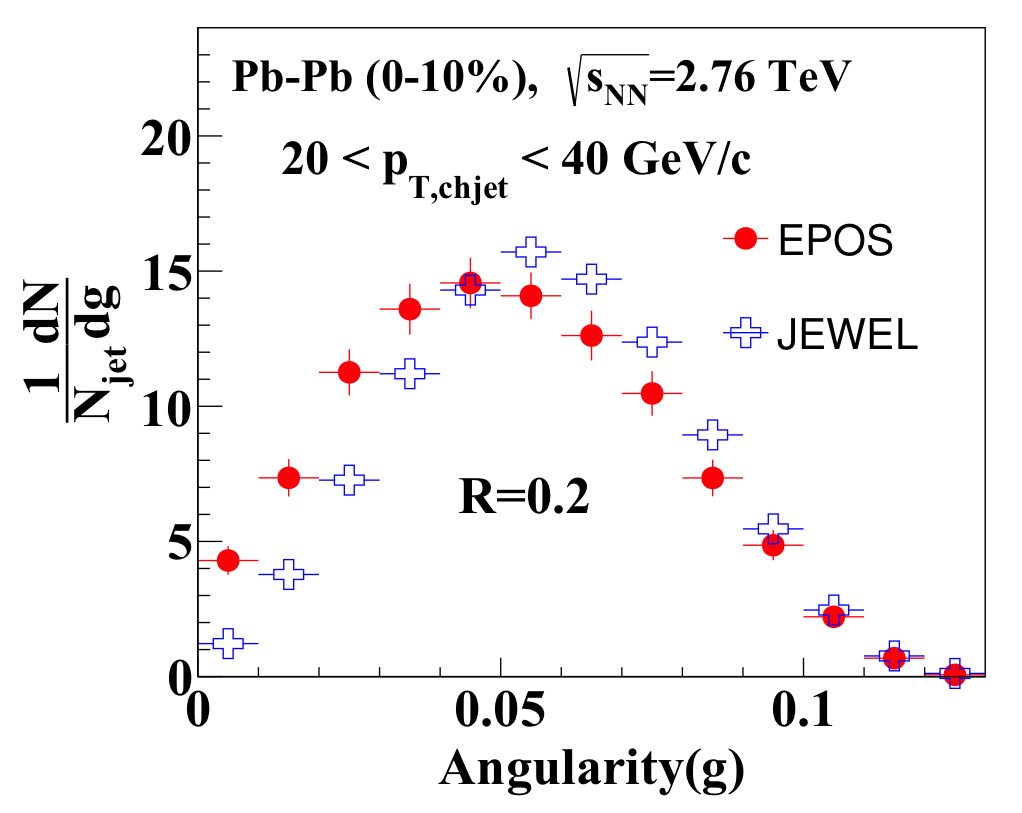}
 	\hfill
 	\includegraphics[width=0.41\textwidth,height=0.38\textwidth]{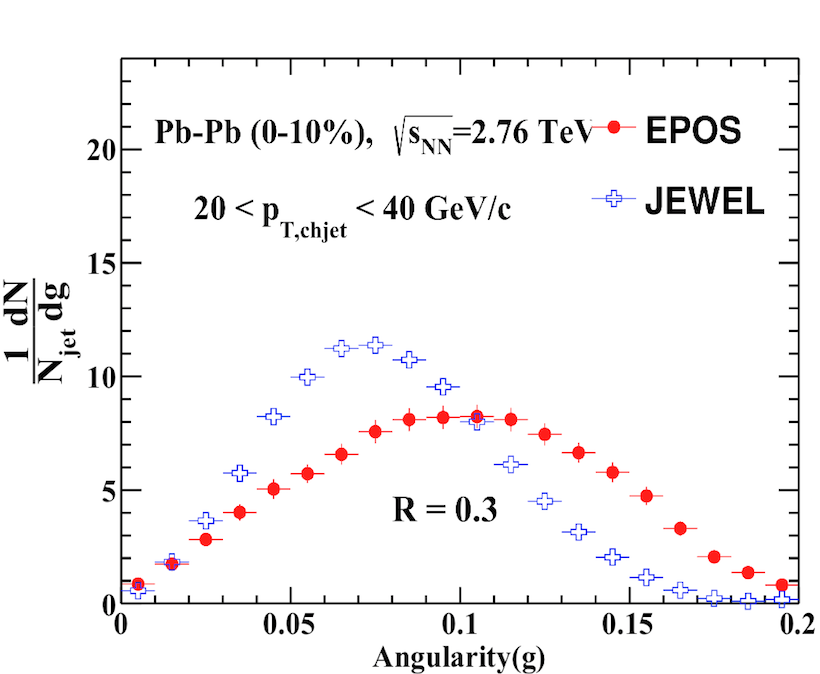}
 	\caption{Angularity (g) measured in 0-10\% central Pb-Pb collisions for inclusive charged jets for R=0.2 and 0.3 using the EPOS-3and JEWEL (recoil OFF) event generators ~\cite{our}.}
 	\label{ref2}
 \end{figure}
 As shown in Fig~\ref{ref2} ~\cite{our}, the jet in EPOS-3 is more collimated than JEWEL for R=0.2, whereas, the jets become broadened due to the medium induced modifications for R = 0.3.
\section{Conclusion}
A comparison between the results from JEWEL recoil off and EPOS-3 shows that JEWEL does not explain the distribution of lost energy at higher radii with respect to the jet-axis, whereas, EPOS-3 explains the effect quite well.

\section*{Acknowledgements}
This material is based upon work supported by the U.S. Department of EnergyOffice of Science, Office of Nuclear Physics under Award Number DE-FG02-92ER-40713. DS would like to thank Federico Ronchetti, Alessandra Fantoniand Valeria Muccifora for their kind help and support throughout this work. Thanks to Klaus Werner for allowingus to use EPOS-3 for this work. Thanks to Dr. Subikash Chaudhury for usefuldiscussions. Thanks to VECC grid computing team for their constant effort tokeep the facility running and helping in EPOS and JEWEL data generation


\paragraph{Funding information}
DS  would  like  to  acknowledge  the  financial  support  from  the  CBM-MUCHproject grant of BI-IFCC/2016/1082(A).



\bibliography{SciPost_Example_BiBTeX_File.bib}

\nolinenumbers

\end{document}